\documentstyle[12pt,psfig]{article}

\newcommand{\be}{\begin{equation}}
\newcommand{\ee}{\end{equation}}
\newcommand{\ba}{\begin{eqnarray}}
\newcommand{\ea}{\end{eqnarray}}

\newcommand{\M}{{\bf M}}		% mixture matrix
\newcommand{\D}{{\bf D}}		% diagonal matrix
\newcommand{\Pn}{{\bf P}}		% permutation
\newcommand{\Mbar}{\bf {\overline{M}}}	% normalized mixture matrix
\newcommand{\mbar}{\overline{M}}	% idem, when used with indices
\newcommand{\J}{{\bf J}}		% synaptic weights
\newcommand{\W}{{\bf W}}		% = J M
\newcommand{\Si}{{\bf S}}		% input signal

\newcommand{\C}{{\bf C}}                % input covariance
\newcommand{\Co}{{\bf C}^0}             % input covariance at \epsilon=0
\newcommand{\Cu}{{\bf C}^1}             % input covariance, first order correction
\newcommand{\ev}{{\bf v}}             	% eigenvectors of \C
\newcommand{\evo}{{\bf v}^0}            % eigenvectors of \Co
\newcommand{\evoT}{{\bf v}^{0T}}        % eigenvectors of \Co, transposed
\newcommand{\s}{{\bf s}}		% sources
\newcommand{\B}{{\bf B}}		% noise covariance
\newcommand{\Nuo}{\mbox{\boldmath $\nu$}_0}	% input noise
\newcommand{\Nu}{\mbox{\boldmath $\nu$}}	% output noise
\newcommand{\h}{{\bf h}}		% = \J\S = psp's
\newcommand{\V}{{\bf V}}		% network output = f(h)
\newcommand{\MT}{{\bf M}^T}		%
\newcommand{\JT}{{\bf J}^T}		%
\newcommand{\NuoT}{\mbox{\boldmath $\nu$}_0^T}	%

\newcommand{\E}{{\cal E} }		% criterion to be minimized
\newcommand{\Idn}{\mbox{$\bf 1$}_N}	% unit matrix NxN
\newcommand{\Idm}{\mbox{$\bf 1$}_m}	% unit matrix mxm

\title{Blind source separation \\in the presence of weak sources}

\author{{\bf\sc J.-P. Nadal} \\
Laboratoire de Physique Statistique de l'ENS\thanks{Laboratoire
associ\'e au CNRS (URA1306), \`a l'ENS et aux Universit\'es Paris 6 et Paris 7.} \\
Ecole Normale Sup\'erieure \\
24, rue Lhomond - 75231 Paris Cedex 05, France,
\\
\\
{\bf\sc E. Korutcheva}{\thanks{Corresponding author: tel: +34-91-398-71-26, 
fax: +34-91-398-66-97, e-mail: elka@fisfun.uned.es;\,\,\,\,\,\,\,\,\,\,\,\,\,
Permanent address: G. Nadjakov Institute of 
Solid State Physics, Bulgarian Academy of Sciences, 1784 Sofia, Bulgaria}} \\ 
Departamento de F\'{\i}sica Fundamental\\
Universidad Nacional de Educaci\'on a Distancia (UNED) \\
c/Senda del Rey No 9  - 28080 Madrid, Spain
\\
\\
and \\ 
\\
{\bf\sc F. Aires} \thanks{Present address: NASA / 
Goddard Institute for Space Studies 
2880 Broadway, New York, NY 10025, USA} \\
Laboratoire de Physique Statistique de l'ENS \\
Ecole Normale Sup\'erieure \\
24, rue Lhomond - 75231 Paris Cedex 05, France
}

\date{}

\begin{document}

\maketitle

\vskip1cm
{\bf Acknowledgments}
\vskip0.5cm

We thank T. Bell for giving us access to his image data base.
This work has been partly supported by the French contract DGA 962557A/DSP.
E.K. warmly thanks for hospitality the Laboratoire de Physique Statistique 
de l'Ecole Normale Sup\'erieure, where this work was performed. 
E.K. has been also supported by the Spanish DGES project PB97-0076 and 
partly by the French-Spanish Program PICASSO and the Bulgarian Scientific 
Foundation Grant F-608.

\newpage

\begin{center}
{\huge Blind source separation \\in the presence of weak sources}
\end{center}

\begin{abstract}
We investigate the information processing of a linear mixture of 
independent sources of different magnitudes. In particular we consider
the case where a number $m$
of the sources can be considered as
``strong'' as compared to the other ones, the ``weak'' sources. We find that 
it is preferable to perform blind source separation
in the space spanned by the strong sources,
and that this can be easily done by first projecting
the signal onto the $m$ largest principal components.
We illustrate the analytical results with numerical simulations.
\end{abstract}


\newpage

\section{Introduction}
During the recent years many studies have been devoted to the 
study of Blind Source Separation (BSS) and more generally
of Independent Component Analysis (ICA) (see e.g. \cite{HJA,cardoso,comon,
amari}).
Within the standard framework one assumes a multidimensional measured signal
to result from a linear mixture of statistically independent components,
or ``sources''. In most cases one makes the optimistic hypotheses that
the number of sources is equal to the dimension of the signal
(the number of captors), and that the unknown mixture matrix is invertible.
The goal of BSS is then to compute an estimate of the inverse of
the mixture matrix in order to extract from the signal the independent components.

In the present paper
we study the effect of having sources with different ``strengths'' when performing
BSS. After giving a proper definition of the strength of a source,
the main purpose of our study is to relate the strength of a source to
its contribution to the information conveyed by the processing
system about the signal, and to consider with more details
the case where some of the sources
are very weak compared to the others.
We will show that in that case it is worthwhile to
project the data onto the space generated by the strong sources
in order to extract meaningful information and to avoid numerical problems.
The contributions to the (projected)
signal from the weak sources can then be considered as noise terms 
added to the linear mixture of strong sources. Since the sources are
independent, this ``noise'' is thus independent of the ``pure'' signal (the part
due to the strong sources).

The paper is organized as follows.
In section \ref{sec:model} we introduce the model and give a precise definition
to the strength of a source.
In section \ref{sec:info} we compute Shannon information quantities from which
we characterize how
each source contributes to the information conveyed by the data and 
by the output of the processing network.
We then discuss the case of a linear mixture of $N$ 
independent sources with $N-m$ ``weak'' sources and $m$ ``strong'' sources.
The results of section \ref{sec:info} show that in such a case it would be
preferable to be able to work in the $m$ dimensional space spanned by
the strong sources.
We show in section \ref{sec:pca} that, with a good approximation,
this is simply done by projecting the data 
onto the $m$ largest principal
components. As a result one can perform BSS in the $m$-dimensional space
where one is dealing with a $m$-dimensional linear mixture corrupted
by a weak input noise.
In section \ref{sec:noisydata} we study,
at first non trivial order in the noise strength,
the expected performance
in the estimation of the $m$ strong sources. Eventually in section
\ref{sec:simul} we present numerical simulations.

\section{The Model}
\label{sec:model}

We consider the information processing 
of a signal which is
a $N$-dimensional linear mixture of $N$ independent sources. At each time $t$
one observes $\Si(t) = \{S_{j}(t), j=1,...,N\}$ which can be written in term
of the unknown sources 
%%\newpage
$\s(t)=\{s_{\alpha}(t), \alpha=1,...,N\}$ as:
\begin{equation}
S_{j}=\sum_{\alpha=1}^{N} M_{j\alpha}\;s_{\alpha}, \hskip2cm j=1,...,N ,
\label{linmix}
\end{equation}
where $\M=\{M_{j\alpha},j=1,...,N,\alpha=1,...,N\}$ is the mixture matrix
assumed to be invertible.
As it is well known, and easily seen from the above equation, it is not possible to
distinguish between the mixture of $\s$ with the matrix $\M$ from
the mixture of $\s'\equiv\Pn\D\s$ with the matrix 
$\M'\equiv\M \D^{-1}\Pn^{-1}$ where $\D$ is an
arbitrary diagonal matrix with non zero diagonal elements, and $\Pn$ an arbitrary
permutation of $N$ indices. If we decide to consider both normalized sources
and normalized mixture matrices, we are left with a diagonal matrix $\D$
which defines the ``strengths'' of the sources. More precisely we write
\begin{equation}
S_{j}=\sum_{\alpha=1}^{N} \mbar_{j\alpha}\;\eta_{\alpha}\;s_{\alpha}, \hskip2cm j=1,...,N
\label{linmixnorm}
\end{equation}
assuming zero mean and unit variance for every source:
\be
<s_{\alpha}>=0, \hskip1cm <s_{\alpha}^2>=1, \hskip1cm \alpha=1,...,N,
\ee
where $<.>$ denotes the average with respect to the (unknown)
sources probability distributions, 
\be
\rho(\s) \;=\; \prod_{\alpha}\;\rho_{\alpha}(s_{\alpha}),
\label{rho}
\ee
and with $\Mbar$ the normalized mixture matrix. The normalization can be chosen
in different ways, and two of them are of particular interest for what follows.
The simplest one is, for each $\alpha$,
\begin{equation}
 \left[\Mbar^T\Mbar\right]_{\alpha\alpha}\;=\;\sum_{j=1}^{N} (\mbar_{j\alpha})^2=1.
\label{norm1}
\end{equation}
The second one is a normalization on the inverse of the mixture matrix:
\begin{equation}
 \left[\Mbar^{-1}\Mbar^{T-1}\right]_{\alpha\alpha}\;=\;\sum_{j=1}^{N} (\left[\mbar^{\;-1}\right]_{\alpha j})^2=1.
\label{norm2}
\end{equation}
Once a particular normalization, such as (\ref{norm1}) or (\ref{norm2}), is chosen, 
the parameters $\eta_{\alpha}$ in (\ref{linmixnorm}) are well defined and
can be understood as the relative strengths of the sources. 

\section{Information processing in the presence of inhomogeneous sources}
\label{sec:info}

Since the mixture matrix is assumed to be invertible, it is in principle possible
to compute an estimate of it. This can be done with any one of the
known blind source separation (BSS) algorithms (see e.g. \cite{comon,cardoso,bellsej,npBlind}).
As a result one obtains an estimate of
the inverse of the mixture matrix, which in our notations can be written as
\be
\frac{1}{\eta_a} \; \left[\mbar^{\;-1}\right]_{a,j} \;\;.
\ee
This shows that it will be dominated by the smallest $\eta$'s, and numerical instabilities or
overflows may occur if some of them are very small.
In many approaches to BSS whitening of the data is first performed. 
The whitened data are then an orthogonal mixture of sources, so that
after this preprocessing one has sources of equal strengths. But this
preprocessing requires a multiplication by the inverse of the eigenvalues,
and this is subject to the same numerical
problems as with the computation of the inverse of the mixture matrix: as we 
will see in section \ref{sec:pca},
small values of $\eta$ leads to the existence of small eigenvalues.

\subsection{Information content of the data}
Let us now compute the amount of information conveyed by the data, $\Si$, about
the sources, that is the mutual information \cite{blahut} $I(\Si,\s)$. To do so we consider
\be
S_{j}=\sum_{\alpha=1}^{N} \mbar_{j\alpha}\eta_{\alpha}s_{\alpha} \;+\; \nu_j,
\hskip2cm j=1,...,N .
\ee
where $\Nu = \{\nu_j, j=1,...,N \}$ is 
a vanishing additive noise, $<\nu_j>=0, <\nu_j\;\nu_k> = b\; \delta_{j,k}$
with $b \rightarrow 0.$ Then
$I(\Si,\s)$ is a constant (that is a quantity that depends on $b$ alone)
plus the data entropy. Since the mixture matrix is invertible, we have
\be
I(\Si,\s) = Const. \;+\;\ln |\det \Mbar| \; +\; \sum_{\alpha} \ln \eta_{\alpha}
\;-\; \sum_{\alpha} \int dh_{\alpha} \rho_{\alpha}(h_{\alpha}) \ln \rho_{\alpha}(h_{\alpha}).
\label{crit1}
\ee
The last term in the above expression is the sum of the source entropies. 
One should remember
that the $s$'s are the 
normalized sources, $<s_{\alpha}^2>=1$. This shows that each source contributes
to the information by a combination of its strength and its entropy: the strength term favors
strong sources, whereas the entropy term favors the sources with a 
probability distribution function (p.d.f.) close to
Gaussian. The entropy terms, however, are bounded:
the entropy of a source cannot exceeds the one of a Gaussian with same
variance, that is
\be 
-\; \int dh_{\alpha} \rho_{\alpha}(h_{\alpha}) \ln \rho_{\alpha}(h_{\alpha}) 
\leq \frac{1}{2} \ln 2 \pi e \;\;.
\ee
Hence the information can be easily dominated by the strength terms, which can be arbitrarily
large.

It is known that for performing BSS perfect knowledge of the sources
distribution is not necessary, and working on the cumulants of order 2 and 3 or 4
is sufficient (see e.g. \cite{comon,npBlind}). We can thus analyze the
result Eq. (\ref{crit1}) by making a close-to-Gaussian 
approximation\cite{comon,npBlind}. If we assume
the sources to have non zero third order cumulants,
\begin{equation}
\lambda_{\alpha}^{(3)}\equiv<s_{\alpha}^{3}>_{c},
\label{3cum}
\end{equation}
we replace the source distribution $\rho_{\alpha}$ by
\be
\hat{\rho}_{\alpha}(s_{\alpha})=\frac{e^{- s_{\alpha}^{2}/2}}{\sqrt{2\pi}}
\left(1+\lambda_{\alpha}^{(3)}\frac{s_{\alpha}(s_{\alpha}^{2}-3)}{6}\right) \;.
\label{closetoG}
\ee
The distribution $\hat{\rho}_{\alpha}$ has the same three first moments
as the true distribution $\rho_{\alpha}$ \cite{abramsteg}.

In the case of a symmetric non-Gaussian distribution, 
the third order cumulants are zero and one has then to
take into account non zero fourth order cumulants.
It is a straightforward exercise to perform the same analysis as below in
that case. For simplicity in this paper we will consider
only the case of non symmetric distributions.

Within this approximation,  Eq.(\ref{closetoG}),   the mutual 
information (\ref{crit1}) 
reads:
\be
I(\Si,\s) = Const. \;+\;\ln |\det \Mbar| \; +\; \sum_{\alpha} \ln \eta_{\alpha}
+ \frac{N}{2} \ln 2 \pi e \;-\; \frac{1}{12} \sum_{\alpha} \; <s_{\alpha}^3>_c^2.
\ee
From the above expression the most important source are those
for which the quantity
\be
<s_{\alpha}^3>_c^2 \;-\; \ln \eta_{\alpha}
\label{crit1bis}
\ee
is the smallest.

We consider now the information that will be conveyed by a network processing the data,
and ask for the contribution to this information by each source when
the network performs BSS.

\subsection{Characterization from infomax}
\label{sec:infomax}

The infomax criterion \cite{linsker,npLownois} will allow us to get some more insight onto the
link between the sources strengths and the amount of information that can be extracted
from the data. 

We consider the information processing of the signal by
a nonlinear network, and we are interested in computing the mutual information
$I(\V,\Si)$ between the input $\Si$ and the output $\V=\{V_i, i=1,...,N\}$ of the network.
Since the signal is  a
linear mixture, the relevant architecture is a linear processing followed by a 
(possibly) nonlinear
transfer function which may differ from neuron to neuron:
\ba
V_i = f_i (h_i) \;+\;\nu_i \\
h_i = \sum_j \; J_{i j} \; (\;S_j\;+\;\nu^0_j\;)\;,
\label{network}
\ea
where $\Nuo = \{\nu^0_j, j=1,...,N \}$ and $\Nu = \{\nu_i, i=1,...,N \}$ 
are additive input and output noise, respectively, with
$<\Nuo>=0, <\Nu>=0, <\nu^0_j\;\nu^0_{j'}> = b^0\; \delta_{j,j'}, 
<\nu_i\;\nu_{i'}> = b\; \delta_{i,i'}$.
The $J_{ij}$ can be viewed as synaptic efficacies and the $h_i$'s
as post-synaptic potentials (PSP).
As explained in the previous section, the noise 
has to be introduced in order to have a nontrivial mutual information,
and we take the limit $0 \leq b^0 << b << 1$.
For strictly zero input noise, $b^0=0$, in the
limit $b\rightarrow 0$ the mutual information
is up to a constant equal to the output entropy.
As shown in \cite{npLownois} its maximization over the choice of
both $\J$ and the transfer functions $f_i$'s leads to BSS. 
One can then derive practical algorithms
for performing BSS \cite{bellsej}. In this limit of $b^0=0$ all the sources 
play the same role, that is the maximum of the mutual information is
independent of the individual sources properties as well as of the
mixture matrix. When one takes into account a non zero input noise,
then at first non trivial order in $\frac{b^0}{b}$ one sees that the
input noise introduces a scale which breaks this invariance.
More precisely, at first order in $\frac{b^0}{b}$
the mutual information $I(\V,\Si)$
can be written (see \cite{npLownois} for details):
\be
I(\V,\Si) \;=\;  I_0(\V,\Si) \;-\; \frac{b^0}{2 b}
\sum_{i=1}^N \Gamma_{ii} \int dh_i \psi_i(h_i) f_i^{'2} \;,
\label{i1}
\ee
where $I_0(\V,\Si)$ is the value at $b^0=0$,
\be
I_0(\V,\Si) \;=\; Const. \;-\; \int d\h \psi(\h) 
\ln \frac{\psi(\h)}{\prod_{i=1}^N f_i'(h_i)}
\ee
and
$\frac{b^0}{b}\;\Gamma_{ii}$
 is the variance of the noise on the PSP $h_i$:
\be
\Gamma_{ii} \equiv \left[\J\JT\right]_{ii} \;.
\ee
Finally, $\psi(\h)$ is the probability distribution of $\h$
induced by the sources input distribution, and
$\psi_i(h_i)$ the marginal distribution of the PSP $h_i$.
At a given $\J$, optimizing with
respect to the choice of transfer functions gives
\begin{equation}
f_{i}'(h_i) = \psi_{i}(h_i)\; \{ \; 1 \;+\; \frac{b^0}{b}\;
\Gamma_{ii} \; \left[ <\psi_{i}^{2}>-\psi_{i}^{2}(h_i) \right] \; \}
\label{fadapted}
\end{equation}
with $<\psi_{i}^{2}> =\int dh_i \psi_i(h_i) \psi_{i}^{2}(h_i) =\int dh_i \psi_i(h_i)^3$.
%%%%For these optimal transfer functions, the mutual information becomes
%%%%\be
%%%%I(\V,\Si)\;=\; Const. \;-\; \int d\h \psi(\h) \ln \frac{\psi(\h)}{\prod_{i=1}^N\psi_{i}(h_i)}
%%%%\;-\; \frac{b^0}{2 b} \sum_{i=1}^N \Gamma_{ii} \int dh_i \psi_i(h_i)^3
%%%%\ee
We now optimize over $\J$. At zeroth order the optimum is reached for
$\J=\M^{-1}$ (up to an arbitrary permutation), so that we write
\be
\W \equiv \J \M = \Idn + \frac{b^0}{b} \W^1 ,
\ee
where $\Idn$ is the $N \times N$ identity matrix.
Expanding the mutual information at first order in $\frac{b^0}{b}$ one
finds that there is no contribution from $\W^1$ to this order.
Hence the mutual information at first order in $\frac{b^0}{b}$
is given by Eq. (\ref{i1}) at $\J=\M^{-1}$, with $f_{i}'$ given
by (\ref{fadapted}) in which we set $\psi_{i}=\rho_i$. This gives
\be
I(\V,\Si) \;=\;  Const. \;-\; \frac{b^0}{2 b}
\sum_{\alpha=1}^N  \Gamma_{\alpha \alpha} \;\int ds_\alpha [\rho_\alpha(s_\alpha)]^3
\ee
with
\be
\Gamma_{\alpha \alpha} \;=\; \left[\M^{-1}\M^{T-1}\right]_{\alpha\alpha} \;.
\ee
One sees that the term depending on $\M$ is what appears in the normalization
(\ref{norm2}) of the mixture matrix. Hence if one chooses this particular
normalization (\ref{norm2}) in order to define the strengths $\eta_{\alpha}$
of the sources, one can rewrite
\be
I(\V,\Si) \;=\;  Const. \;-\; \frac{b^0}{2 b}
\sum_{\alpha=1}^N  \; \frac{1}{\eta_{\alpha}^2} \; <\rho_{\alpha}^2>
\label{i2}
\ee
with $<\rho_{\alpha}^2>=\int ds_\alpha \left[\rho_\alpha(s_\alpha)\right]^3$.
The above expression shows how each source $\alpha$ contributes to the mutual
information in term of its strength $\eta_{\alpha}$ and its pdf $\rho_{\alpha}$.

Within the close-to-Gaussian approximation (\ref{closetoG}) one gets
\be
I(\V,\Si) \;=\;Const \;-\;\frac{b^0}{b} 
\sum_{\alpha=1}^{N} \; <s_{\alpha}^3>_c^{2} \; \frac{1}{\eta_{\alpha}^2}.
%%%\sum_{\alpha=1}^{N} \lambda_{\alpha}^{(3)^{2}} \Gamma_{\alpha\alpha}.
\label{i3}
\ee
Hence the sources which contribute the most to the conveyed information
are those for which the quantity
\begin{equation}
\E_{\alpha} \equiv <s_{\alpha}^{3}>_{c}^{2}\;\frac{1}{\eta_{\alpha}^{2}} 
\label{crit2}
\end{equation}
is the smallest. One should remember that, here, $\eta_{\alpha}$
is given by
\be
\frac{1}{\eta_{\alpha}^2} = \sum_{j=1}^{N} \left(\left[\M^{\;-1}\right]_{\alpha j}\right)^2.
\label{str2}
\ee

\subsection{Discussion}
As already seen when computing the mutual information
between the data and the sources, a source will contribute
if it is strong and/or close to Gaussian. However the particular combination which
appears here is different from the one we obtained in the previous section:
here we have a multiplicative combination of strength and cumulant, whereas
in (\ref{crit1bis}) it was an additive combination.

An important practical remark is that,
if the third order cumulants are zero, the close-to-Gaussian approximation
has to take into account the fourth order cumulants. Then,
instead of (\ref{crit1bis}) and (\ref{crit2})
one gets similar expressions with the fourth order cumulants in place of the
third order ones.

The criterion (\ref{crit2}) can be used in
different ways, depending on the particular application considered.
The quantity $\E_{\alpha}$ is zero for Gaussian sources, whatever their
strengths. This is not surprising since the Shannon information
is maximal for Gaussian distributions. However in many cases
the Gaussian part of the signal is considered as ``noise'', and the non Gaussian
part is the ``meaningful'' part, the ``true'' signal. Hence mutual information
can be used as a cost function in order to extract this noise, in particular
when it is strong, which can then
be subtracted from the input signal. In cases where one has distributions
of similar shapes, (\ref{crit2}) suggests to use the strength as defined
in (\ref{str2}) to order the sources and select the most relevant ones.

To conclude the present section \ref{sec:info}, we see that the intuitive
idea that weak sources can be considered as noise terms and cannot be
estimated, can be quantified from various point of views. From the purely numerical
aspect, the mixture matrix is close to be singular; the information content of
the data, the amount of information conveyed by a processing channel, are
seriously diminished by the presence of weak sources.
From this analysis, it appears clearly that it would be preferable to
be able to project the data onto the space spanned by the strong sources,
in order to work in a space of smaller dimension with sources of
similar strengths. In the next section we show that this
is simply done by making use of the principal component analysis.

\section{Principal Component Analysis}
\label{sec:pca}

A standard approach in data processing consists in first performing the
principal component analysis (PCA), and then projecting the data 
onto the eigenspace associated with the largest eigenvalues. 
In the present 
context of BSS, 
it is reasonable to expect the space spanned by the strong sources
to be essentially the same as the one associated to the largest principal components.
It is the purpose of this section to give a positive and more
precise  answer to this question.

%%% As a first remark we note the following 
%%% link between the strengths $\eta_{\alpha}$ and the principal components.
%%% Suppose one has obtained an estimate $\J$ of the inverse
%%% of the mixture matrix. In order to compare with the model (\ref{linmix}),
%%% we have first to normalize $\J$ so that the source estimates, $\J \Si$,
%%% have unit variance. Hence we replace $\J$ by 
%%% $\J_{\alpha,j}/\sqrt{[\J \C \J^T]_{\alpha,\alpha}}$
%%% where $\C$ is the covariance matrix of the data.
%%% The estimate of the mixture matrix is then the inverse of this matrix, and
%%% the strengths defined by (\ref{norm2}) are, in term of the first matrix $\J$,
%%% \be
%%% \eta_\alpha^2 = \frac{[\J \C \J^T]_{\alpha,\alpha}}{[\J\J^T]_{\alpha,\alpha}}
%%% \ee
%%% This shows the dependency of the $\eta$'s in the covariance matrix of the data,
%%% and the independence in the global normalization of $\J$.
%%% 
We consider the specific 
case where $m$ source are strong, while $N-m$ sources are weak.
More precisely, choosing for later convenience the normalization (\ref{norm1}),
we assume
\ba
\eta_{\alpha}\sim O(1=\epsilon^0) \;\mbox{for}\;\alpha=1,...,m \nonumber \\
\eta_{\alpha}\sim O(\epsilon)\;\mbox{for}\;\alpha=m+1,...,N ,
\ea
where $\epsilon$ is a small parameter, $\epsilon <<1$. This is equivalent to 
state that there is a gap in the spectrum of eigenvalues at the $\lambda_m$,
with $\lambda_{m+1} << \lambda_m$.

We assume that the reduced $N \times m$ mixture matrix $M^0$, 
$\{M^0_{j\alpha}=M_{j\alpha}, j=1,...,N;\alpha=1,...,m\}$ is of rank $m$, so that
the ($N \times N$) correlation matrix (the covariance of the input signal)
 $\Co$, which would be obtained at $\epsilon \equiv0$,
has $m$ non zero eigenvalues.
It is a standard exercise in perturbation theory 
\cite{perturbtheory}
to study the behavior of the eigenvalues and eigenvectors of a symmetric matrix,
here the covariance matrix $\C$ of the inputs,
at first non trivial order in the small parameter $\epsilon$. 
The eigenvalues have a smooth behavior with $\epsilon$: the $m$ largest eigenvalues
of $\C$ are, at first non trivial order, the $m$ non zero eigenvalues of $\Co$ shifted by
quantities of order $\epsilon^2$,
and the $N-m$ smallest ones are of order $\epsilon^2$. However the eigenvectors
are very sensitive to small variations of $\epsilon$ - this is related to the fact that
the mixture matrix $M$ is closed to be singular for small $\epsilon$. More precisely, 
one gets the following results.

One can write $\C$ as
\begin{equation}
\C = \Co + \epsilon^2 \Cu ,
\label{c0+c1}
\end{equation}
where $\Co$ is the correlation of the inputs that would be obtained
without the weak sources ($\epsilon\equiv0$),
and $\epsilon^2 \Cu$ contains all the contributions from
the weak sources.
We denote by
$\lambda_{\alpha}^0$ the
eigenvalues of $\Co$, with 
$\{ \lambda_{\alpha}^0, \alpha=1,...,m\}$ non zero
and $\lambda_{\alpha}^0=0$ for $\alpha=m+1,...,N$.
The associated eigenvectors
$\{\evo_{\alpha}, \alpha=1,...,N\}$ form an orthonormal basis. 
If all the eigenvalues of $\Co$ are different (hence in particular
$N=m+1$), then, at first order, the eigenvalues of $\C$ are
\ba
\lambda_{\alpha} \;=\; \lambda_{\alpha}^0 \;+\; \epsilon^2 \; \lambda_{\alpha}^1 \nonumber \\
\lambda_{\alpha}^1 \;=\; \evoT_{\alpha} \Cu \evo_{\alpha}
\;\;\;(\alpha=1,...,N),
\label{ev1}
\ea
and the corresponding eigenvectors are
\be
\ev_{\alpha} \;=\; \evo_{\alpha}  \;+\; \epsilon^2 \; 
\sum_{\beta\neq\alpha} \evo_{\beta} \; \frac{\evoT_{\alpha} \Cu \evo_{\beta}}{\lambda^0_{\alpha}-\lambda^0_{\beta}}
\;\;\;(\alpha=1,...,N).
\label{ev2}
\ee
If there are degenerate eigenvalues  (in particular the null eigenvalue
is degenerate for $N>m+1$), this is modified
as follows. Suppose $\Co$ has only $r<N$ different eigenvalues, $\mu_1 > \mu_2 >... > \mu_r$,
with degeneracies $q_a, a=1,...,r$ ($\sum_a q_a = N$, $\mu_r=0$ if $N>m+1$). We have
\be 
\lambda_{\alpha}^0 =\mu_a \;\mbox{for} \;\; \sum_{b=1}^{a-1}q_b < \alpha \leq \sum_{b=1}^{a} q_b \;\equiv\;\alpha_a
\ee
and we set $\alpha_0 \equiv0$.
Consider an eigenvalue $\mu_a$ with degeneracy $q_a >1$. The eigenvectors
of $\Co$ associated to $\mu_a$, $\{\evo_{\alpha}, \alpha_{a-1}<\alpha\leq\alpha_{a}\}$,
 form an orthonormal basis of this eigenspace of dimension $q_a$, and this base is defined
up to an arbitrary orthogonal transformation. This arbitrariness is removed at first
non trivial order in $\epsilon$, together with the removal of the eigenvalue degeneracy:
the new $q_a$ eigenvalues for $\{\alpha_{a-1}<\alpha\leq\alpha_{a}\}$
are given by Eq. (\ref{ev1}), where the $\evo$'s form the particular
$q_a \times q_a$ orthogonal matrix which diagonalizes $\Cu_a$, the restriction of
the matrix $\Cu$ to the eigenspace of $\mu_a$,
the $\lambda_{\alpha}^1$ being then the eigenvalues of $\Cu_a$.

The eigenvectors $\ev$ are now given by an equation similar to (\ref{ev2}), with
the sum over $\beta\neq\alpha$ replaced by a sum 
over the $\beta$ such that $\lambda_{\beta} \neq \lambda_{\alpha}$,
and a new term specific to each degenerate eigenvalue $\mu_a$:
\ba
\ev_{\alpha} \;=\; \evo_{\alpha}  \;+\; \epsilon^2 \; 
\sum_{\beta:\lambda_{\beta} \neq\lambda_{\alpha}} \evo_{\beta} \; \frac{\evoT_{\alpha} \Cu \evo_{\beta}}{\lambda^0_{\alpha}-\lambda^0_{\beta}}
\nonumber \\
+\; \epsilon^2 \;\sum_{\beta:\lambda_{\beta} =\lambda_{\alpha}} \;X_{\alpha,\beta} \;\evo_{\beta} \;\;\;(\alpha=1,...,N)\;,
\label{ev3}
\ea
where the $\evo$ are chosen as just explained, 
and $X_{\alpha,\beta}$ is an arbitrary antisymmetric matrix.

The final result is thus that the space generated by the $m$ eigenvectors associated
to the $m$ largest eigenvalues is, to order $\epsilon^2$, the same space
as the one which would be obtained in the absence of the weak sources.
Projecting the data onto this space is then equivalent to working with the 
$m$-dimensional signal which is the mixture
of the $m$ strong sources, weakly corrupted by an additive noise.

\section{BSS with noisy data}
\label{sec:noisydata}

Let us now assume that we have pre-processed the data by 
projecting it onto the $m$ largest principal components.
To avoid the introduction of a new notation, in the following
$\{S_j,j=1,...,m\}$ will denote these preprocessed data (projections)
instead of the data themselves.  
Instead of the model Eq.(\ref{linmix}) we have thus to consider the model
\begin{equation}
S_{j}=\sum_{\alpha=1}^{m} M_{j\alpha}s_\alpha \;+\; \nu^0_j ,
\hskip3cm j=1,...,m .
\label{linnoise}
\end{equation}
The matrix  $\M$ is now a $m \times m$ invertible mixture matrix, 
such that $\M\MT$ has $m$ non zero,
of order $1=\epsilon^0$, eigenvalues. The $s_\alpha$'s ($\alpha=1,...,m$)
are the sources of interest, and the 
$\nu^0_j$'s
are additive noises, resulting from the weak sources, as explained in
the previous section. This noise 
$\Nuo = \{\nu^0_j, j=1,...,m \}$ 
is uncorrelated with the $m$ (strong) sources,
and of arbitrary distribution $P(\Nuo)$. Since we are working in the small $\epsilon$ regime, 
all we will need is to characterize this distribution by its first two cumulants:
\begin{eqnarray}
\label{noise}
<\Nuo>=0
\nonumber\\
<\Nuo \NuoT>=\epsilon^2 \; \B \;,
\end{eqnarray}
where $\B$ is a (possibly non diagonal) $m \times m$ symmetric matrix.
The problem we are considering now
is thus strictly the same as the one of performing BSS on a linear
mixture of $m$ sources corrupted by some additive input noise, which,
although small, cannot be neglected.

\subsection{The Mutual Information}
\label{sec:MI}

In this section we consider this noisy BSS problem within the infomax approach 
as formulated in \cite{npLownois}. The network we consider has the same architecture
as the one defined in Eq. (\ref{network}), but with $m$ inputs and outputs: 
\ba
V_i = f_i (h_i) \;+\;\nu_i \\
h_i = \sum_{j=1}^{m} \; J_{i j} \; (\;S_j\;+\; \nu^0_j\;)\;\;\;\;i=1,...,m,
\label{network-m}
\ea
with $<\nu_i\;\nu_{i'}> = b\; \delta_{i,i'}$.
The limit to be considered here is the one of a vanishing output noise,
$b \rightarrow 0$, but at a given input noise level:
\be
0 < b << \epsilon^2.
\ee
Another important difference with the calculation done in section \ref{sec:infomax},
is that here we are interested in computing the information conveyed about the global
input, $\Si + \Nuo$, and not about the ``pure'' signal alone $\Si$.
Indeed, in section \ref{sec:infomax} we considered some input noise
corresponding to some noise at the level of the receptors, whereas
here the actual signal is the global input, $\Si + \Nuo$, in which
{\em we} have decided to call ``(pure) signal'' the part coming
from the strong sources and ``noise''
the part due to the weak sources.

In this limit of vanishing output noise, the mutual information  $I(\V,\Si+\Nuo)$
between the output and the input of the network is up to a constant equal to
the output entropy.
To simplify the analysis, we assume a full adaptation of the transfer functions,
which means\cite{npLownois}, for $\J$ given,
\be
f'_{i}(h_i)=\psi_i(h_i), i=1,...,m \;,
\ee
where $\psi_i(h_i)$ is the marginal probability distribution of the PSP $h_i$.
As a result the mutual information is up to a constant
equal to the redundancy between the PSP's \cite{npLownois}:
\begin{equation}
I(\V,\Si)=Const. -\int d^m\h \psi(\h) 
\ln \frac{\psi(\h)}{\prod_{i=1}^{m}\psi_i(h_i)} .
\label{MI1}
\end{equation}

\subsection{Maximization in the small $\epsilon$ limit}

In term of the sources distributions, the distribution $\psi(\h)$ is given by:
\begin{equation}
\psi(\h)=\int\prod_{\alpha=1}^{m} ds_{\alpha} \;\rho_{\alpha}(s_{\alpha})
\; \int d^m\Nuo P(\Nuo)
\; \prod_{i=1}^{m} \; \delta(h_i-\sum_{\alpha} [JM]_{i\alpha}\;s_{\alpha} \;-\; \sum_j J_{ij}\;\nu^0_j) .
\label{psi}
\end{equation}
Since in Eq. (\ref{psi}) the noises $\nu^0_j$ are $\sim O(\epsilon)$
we can perform an expansion,
leading to the following expression:
\be
\psi(\h)= \{ \;1\;+\; \frac{\epsilon^2}{2}\; 
\sum_{i,i'} \left[\J\B\JT\right]_{ii'}\; \partial_i \partial_{i'} \; \} \;\psi^{0}(\h) \;,
\label{psi2}
\ee
where $\partial_i$ means the partial derivative with respect to $h_i$, and
$\psi^{0}(\h)$ is the p.d.f. that would be obtained at $\epsilon=0$.
Because the noise has zero mean there is no term of order $\epsilon$ in (\ref{psi2}).

%%%%\begin{equation}
%%%%\psi(\h)=\psi^{0}(\h) + \epsilon^2\; [\W\B\WT]_{ii} 
%%%%\frac{d^2}{dh_{i}^{2}}\psi_{i}^{0}(h_{i}^{0})+\epsilon^2\sum_{i\neq i'}(WBW^{T})_{ii'}
%%%%\frac{d}{dh_i}\frac{d}{dh_{i'}}\psi_{i}^{0}(h_{i}^{0}) ,
%%%%\label{psiexpand}
%%%%\end{equation}
%%%%where we have already used an expansion of $\psi_i(h)$ on the powers of $\epsilon$:
%%%%\begin{equation}
%%%%\label{series}
%%%%\psi_i(h_i)=\psi_{i}^{(0)}(h_i)+ \epsilon \psi_i^{(1)}(h_i) 
%%%%+ \epsilon^2 \psi_{i}^{(2)}(h_i)
%%%%\end{equation}

We consider now the maximization of the mutual information over the choice of $\J$,
taking into account that $\epsilon$ is small.
If $\epsilon$ was strictly zero, we would be back to the noiseless BSS
problem for which the optimum is reached for
$\J=\M^{-1}$ (up to an arbitrary permutation). So for nonzero $\epsilon$
we write
\be
\W \equiv \J \M = \Idm \;+\; \epsilon\; \W^1 \;+\; O(\epsilon^2) \;,
\label{ww}
\ee
where $\Idm$ is the $m \times m$ identity matrix, and the correction
is a matrix of order at least $\epsilon$. Since $\W$ depends now on $\epsilon$
we can also expand $\psi^{0}$ in powers of $\epsilon$,
and finally $\psi(\h)$ can then be written as
\be
\psi(\h) = \left[\prod_{\alpha} \rho_{\alpha}(h_{\alpha})\right] \; 
\left[ \;1 \; + \; \epsilon \; Q[\h]
+ R[\h] \; \right]
\label{rr}
\ee
with
\be
Q[\h] \equiv - \; 
\sum_{\alpha, \beta} [\ln \rho_{\alpha}]' \; W^1_{\alpha \beta} h_\beta
\;- Tr \W^1 \;
\label{q}
\ee
and $R[\h]$ contains terms of order at least $\epsilon^2$, coming from both
$\W$, equation (\ref{ww}), and $\B$, equation (\ref{psi2}).
Similarly, for the marginal distributions:
\be
\psi_\alpha(h_\alpha) = \rho_{\alpha}(h_{\alpha}) \; \{ 1 
+ \epsilon \; Q_{\alpha}[h_\alpha]
+ R_{\alpha}[h_\alpha] \; \}\;,
\label{rra}
\ee
with
\be
Q_{\alpha}[h_\alpha] \equiv 
-\; [\ln \rho_{\alpha}]' W^1_{\alpha \alpha} h_\alpha - W^1_{\alpha \alpha} \;.
\label{qa}
\ee
The substitution of Eq.(\ref{rr}) and (\ref{rra}) in the expression (\ref{MI1}) gives 
then for the mutual information, at first non trivial order:

\begin{eqnarray}
\label{MI2} 
I(\V,\Si) & = & I_{0}(\V,\Si) \;-\;
\frac{\epsilon^2}{2} \; \int\prod_{\alpha=1}^{m} dh_{\alpha} \;\rho_{\alpha}(h_{\alpha})
\; \left[ Q[\h] - \sum_{\alpha} Q_{\alpha}[h_\alpha] \right]^2 .
%%%\frac{\epsilon^{2}}{2}\{\sum_{i,i'}B_{ii'}
%%%\int\h\frac{\partial^{2}\psi^{0}}{\partial h_i\partial h_i'} \ln \psi^{0} - 
%%%\nonumber\\
%%%& & \sum_{i}B_{ii}
%%%\int dh_i \frac{\partial^{2} \psi_{i}^{0}}{\partial h_{i}^{2}} \ln \rho_{i} \} .
%%%%% I(\V,\Si) & = & I_{0}(\V,\Si)-
%%%%% \frac{\epsilon^{2}}{2}\{\sum_{i,i'}(WBW^T)_{ii'}
%%%%% \int\h\frac{\partial^{2}\psi^{0}}{\partial h_i\partial h_i'} \ln \psi^{0} - 
%%%%% \nonumber\\
%%%%% & & \sum_{i}(WBW^T)_{ii}
%%%%% \int dh_i \frac{\partial^{2} \psi_{i}^{0}}{\partial h_{i}^{2}} \ln f'_{i} \} .
\end{eqnarray}
%%%with here $\psi^{0}(\h) = \prod_i \psi_{i}^{0}(h_i) = \prod_i \rho_i(h_i)$.
The term $I_{0}(\V,\Si)$ corresponds to the part of the mutual information which does 
not take into account the weak sources. It is the same as if one computes the 
mutual information between the output $\V$ and the signal $\M\s$; 
$I(\V,\M\s)$. 
The fact that there is no term of order $\epsilon$ in (\ref{MI2}) 
can be understood as coming
from the normalization conditions $\int d\h \psi^{0}(\h) =1$ and 
$\int dh_\alpha \psi^{0}_\alpha=1$, which imply
$$
\int\prod_{\alpha=1}^{m} dh_{\alpha} \;\rho_{\alpha}(h_{\alpha})
\;\;Q[\h] = 0
$$
and
$$
\int dh_{\alpha} \;\rho_{\alpha}(h_{\alpha}) \; Q_{\alpha}[h_\alpha]=0
$$
(these properties can be easily checked by performing 
the integrations using the explicit expressions (\ref{q}) and (\ref{qa})).
One has similar properties for the quantities
of order $\epsilon^2$, $R[\h]$ and $R_{\alpha}[h_\alpha]$ defined in (\ref{rr})
and (\ref{rra}), so that they do not contribute at this order $\epsilon^2$
in the final result (\ref{MI2}).

Now one has
\be
Q[\h] - \sum_{\alpha} Q_{\alpha}[h_\alpha] = -\;\sum_{\alpha \neq \beta} 
[\ln \rho_{\alpha}]' \; W^1_{\alpha \beta} h_\beta \;.
\ee
The mutual information is maximized when the quadratic term in (\ref{MI2})
is minimized, that is for $W^1_{\alpha \beta}=0$ for $\alpha \neq \beta$.
It follows that there is 
no correction to the mutual information at order $\epsilon^2$ and that corrections due to the 
weak sources appear at order $\epsilon^4$.

\section{Numerical simulations}
\label{sec:simul}

In this section we illustrate our analysis by numerical simulations.
We test the above analysis on the following toy example. We consider
the ICA of natural images performed in \cite{bellsej}.
First we reproduce the results
in \cite{bellsej}  (not shown here).
We then create a new data base with artificially increased
component strengths: new images are computed as a linear
mixture of the previous ICA basis function but the 
strength of 20 components was augmented 100 times compared
to the other 124.
We performed ICA in this new data base, with the same
algorithm based on infomax \cite{npLownois,bellsej}, but after
projecting the data onto the 20 largest principal components.
The resulting basis function represented on the Figure 1
shows the efficiency of PCA preprocessing: we find the good
20 stronger components and the computational time is considerably decreased.

For such a signal, the PCA analysis is identical to a Fourier
analysis, and therefore dropping the smallest eigenvalues means neglecting
high frequencies. One thus expect to extract components
which are smoothed versions of components extracted when working
with the full space. This is indeed the case as shown on Figure 1.

\section{Concluding remarks}

We have discussed the task of Blind Source Separation
in the case of a mixture of sources of unequal strengths.

We have presented different, but related, ways
of defining the relative strengths of the sources.
In particular,
when non zero input noise is taken into account
the contribution of a source to the conveyed information
can be characterized by a criterion
which combines the mixture matrix elements and the third cumulant of the 
source distribution. This allows to define the strength of a source
once a proper normalization of the mixture matrix is assumed.
Conversely, this study shows which sources will be ``preferred''
by the infomax criterion (which part of the signal is
more likely to be well extracted by an ICA performed with infomax).

The analysis indicates also that, although arbitrary, the assumed normalization
of the mixture matrix may have an important practical
role in the analysis of the outcome of an ICA,
whenever one wants to extract the ``meaningful'' sources.
Which part of the signal is more important is of course
an application dependent notion. Prior knowledge related to
a given case should allow to define the proper normalization
from which the appropriate scale of source strengths can be defined.
Conversely each chosen normalization implies a particular physical 
interpretation which should be kept in mind when analyzing
the outcome of an ICA.

We have considered with more details the particular case
of the information processing of a linear mixture of independent 
sources when some of them are very weak 
as compared to the other sources. One should note that
in such case the notion
of strong {\em versus} weak is independent of the mixture matrix normalization.
It is easily seen that
the presence of weak sources leads to an almost singular mixture matrix, and
this manifests itself by the existence
of very small eigenvalues in the PCA analysis.
We have shown that it is relevant to
project the input data onto the largest principal
components in order to extract the strongest independent sources.
We have thus quantified the intuitive idea that the subspace, where
most of the data live, is mainly spanned by the strongest independent
sources.
We illustrated this result on the ICA of the image data base
studied in \cite{bellsej}.

A possible situation where the PCA will not be (sufficiently) helpful
is when the strong sources generate a linear space of dimension smaller  
than the number of sources. This space will be found by the PCA.
After projection onto the largest PC's, one has then to deal  
with an 
ICA with a number of sources
larger than the number of captors. This is an interesting problem which
has received 
considerable attention recently, and several algorithms have been
proposed.
Our analysis suggests then that it can be meaningfull to project onto
the largest PC's (in order to eliminate the weak sources) and yet to
search
for a number of (strong) IC's larger than the number of largest PC's.

\newpage

%%\bibliography{perso}

\begin{thebibliography}{1}

\bibitem{abramsteg}
M.~Abramowitz and I.~A. Stegun 1972,
\newblock {\em Handbook of Mathematical Functions}
\newblock (Dover).

\bibitem{amari}
S.-I. Amari and J.-F. Cardoso 1997,
\newblock Blind source separation-semiparametric statistical approach,
\newblock {\em IEEE Tr. on Signal Processing}, 45(11):2692, Special issue on
neural networks.

\bibitem{bellsej}
A.~Bell and T.~Sejnowski 1995,
\newblock An information-maximization approach to blind separation and blind
  deconvolution.
\newblock {\em Neural Computation}, 7:1129--1159.

\bibitem{blahut}
R. E. Blahut 1988,
\newblock {\em Principles and Practice of Information Theory}
\newblock (Addison-Wesley, Cambridge MA).

\bibitem{cardoso}
J.-F. Cardoso 1989,
\newblock Source separation using higher-order moments,
\newblock in {\em Proc. Internat. Conf. Acoust. Speech Signal Process.,
Glasgow 1989}, pp. 2109--2112.

\bibitem{comon}
P.~Comon 1994,
\newblock Independent component analysis, a new concept ?
\newblock {\em Signal Processing}, 36:287--314.

\bibitem{HJA}
J. Herault, C. Jutten and B. Ans 1985,
\newblock D\'etection de grandeurs primitives dans un message composite
par une architecture de calcul neuromim\'etique en apprentissage non
supervis\'e,
\newblock in {\em Proc. GRETSI, Nice 1985} pp. 1017-1020.

\bibitem{linsker}
R.~Linsker 1988,
\newblock Self-organization in a perceptual network.
\newblock {\em Computer}, 21:105--17.

\bibitem{perturbtheory}
%%Perturbation Theory: Any good ref. ?
A. Messiah 1961,
\newblock {\em Quantum mechanics} (Elsevier Science Publishers, Amsterdam).

\bibitem{npLownois}
J.-P. Nadal and N.~Parga 1994,
\newblock Nonlinear neurons in the low-noise limit: a factorial code maximizes
  information transfer
\newblock {\em Network: Computation in Neural Systems}, 5:565--581.

\bibitem{npBlind}
J.-P. Nadal and N.~Parga 1997,
\newblock Redundancy reduction and independent component analysis: Algebraic
  and adaptive approaches
\newblock {\em Neural Computation}, 9(7):1421--1456.

\end{thebibliography}
\bibliographystyle{plain}

\newpage
{\bf Figure Captions}
\vskip1cm
Figure 1. Basis functions of the ICA solution

\newpage

\begin{figure}[htbp]
 \begin{center}
 \leavevmode
 \centerline{\hbox{
 \psfig{figure=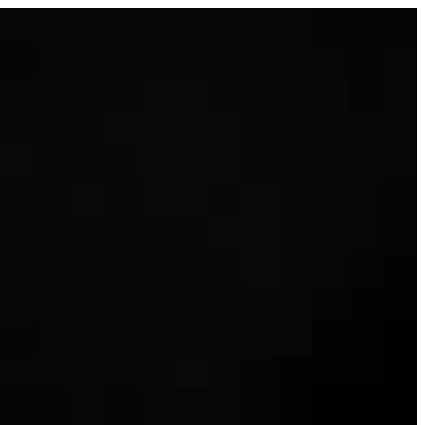,height=2.0cm}
 \psfig{figure=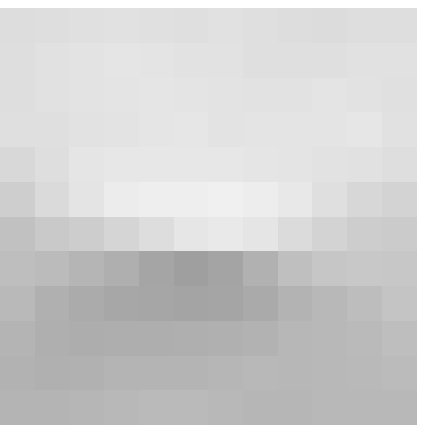,height=2.0cm}
 \psfig{figure=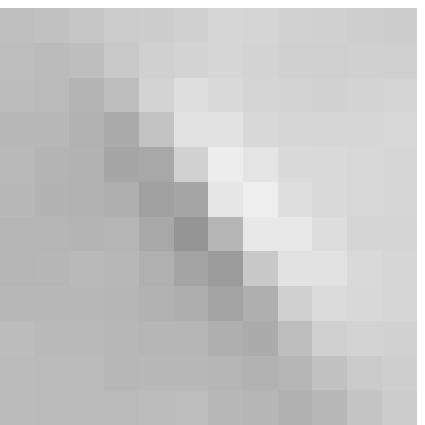,height=2.0cm}
 \psfig{figure=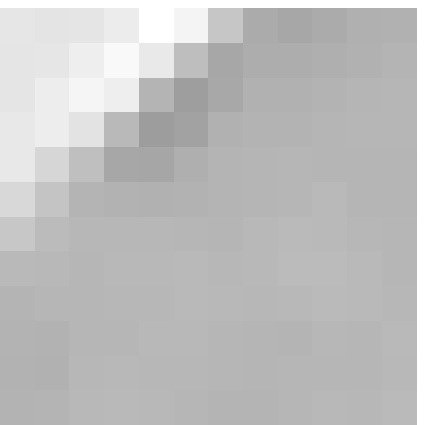,height=2.0cm}
 \psfig{figure=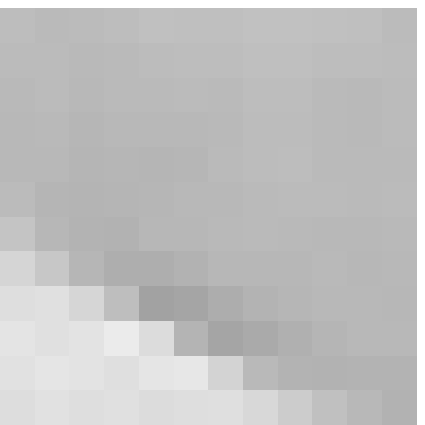,height=2.0cm}
 }}
 \end{center}
 ~\hspace{2.4cm} Bas~ft 0001 \hspace{0.0cm} Bas~ft 0002 \hspace{0.0cm} Bas~ft 0003
 \hspace{0.0cm} Bas~ft 0004 \hspace{0.0cm} Bas~ft 0005
 \begin{center}
 \leavevmode
 \centerline{\hbox{
 \psfig{figure=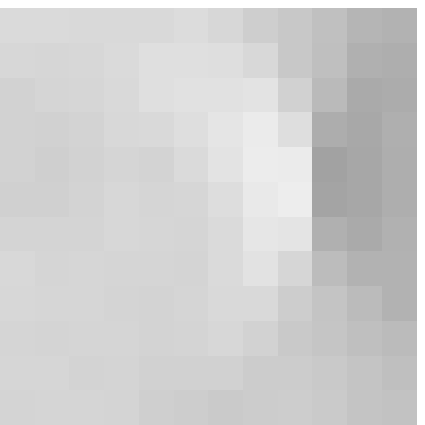,height=2.0cm}
 \psfig{figure=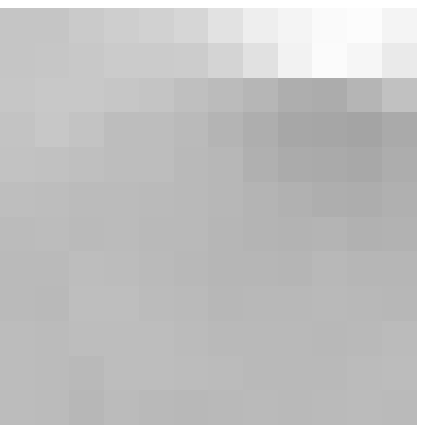,height=2.0cm}
 \psfig{figure=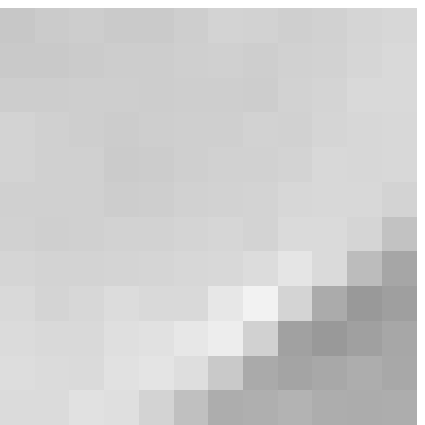,height=2.0cm}
 \psfig{figure=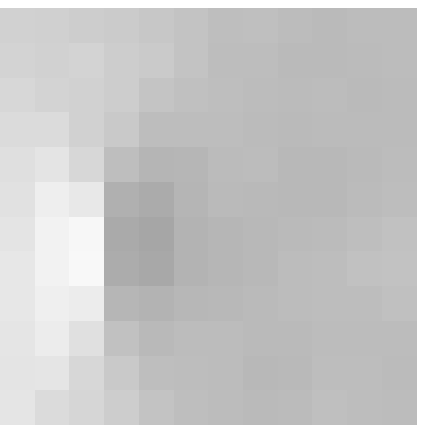,height=2.0cm}
 \psfig{figure=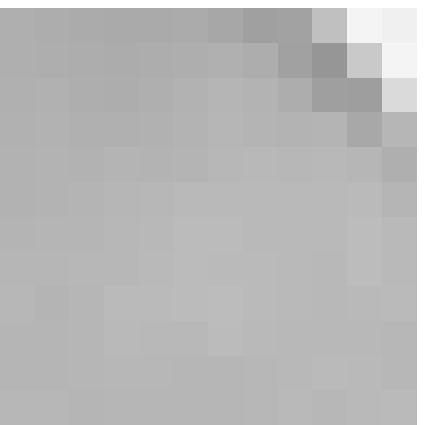,height=2.0cm}
 }}
 \end{center}
 ~\hspace{2.4cm} Bas~ft 0006 \hspace{0.0cm} Bas~ft 0007 \hspace{0.0cm} Bas~ft 0008
 \hspace{0.0cm} Bas~ft 0009 \hspace{0.0cm} Bas~ft 0010
 \begin{center}
 \leavevmode
 \centerline{\hbox{
 \psfig{figure=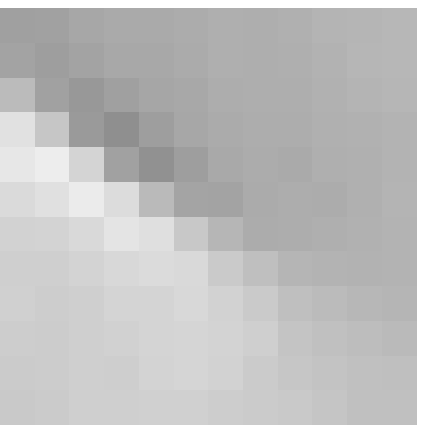,height=2.0cm}
 \psfig{figure=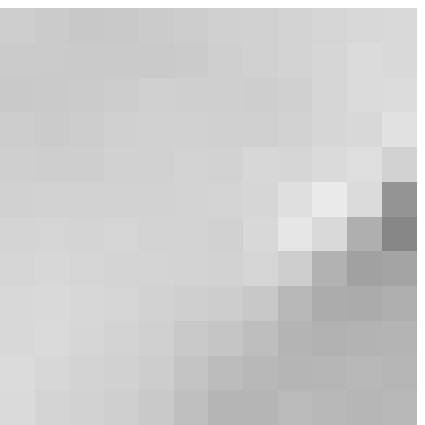,height=2.0cm}
 \psfig{figure=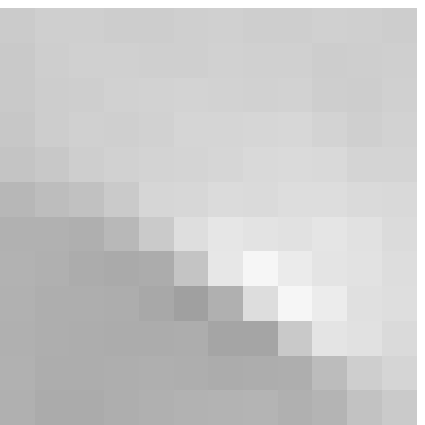,height=2.0cm}
 \psfig{figure=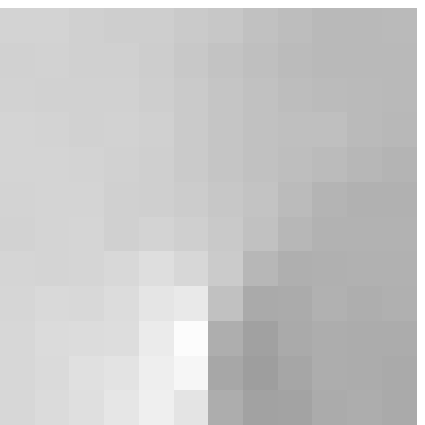,height=2.0cm}
 \psfig{figure=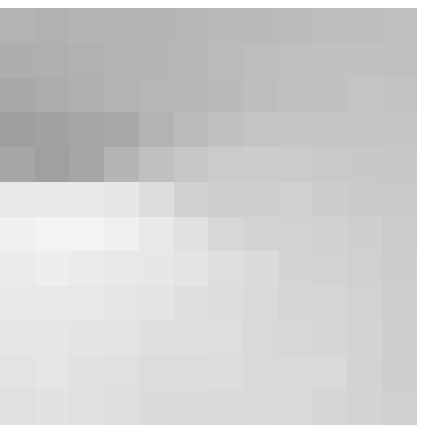,height=2.0cm}
 }}
 \end{center}
 ~\hspace{2.4cm} Bas~ft 0011 \hspace{0.0cm} Bas~ft 0012 \hspace{0.0cm} Bas~ft 0013
 \hspace{0.0cm} Bas~ft 0014 \hspace{0.0cm} Bas~ft 0015
 \begin{center}
 \leavevmode
 \centerline{\hbox{
 \psfig{figure=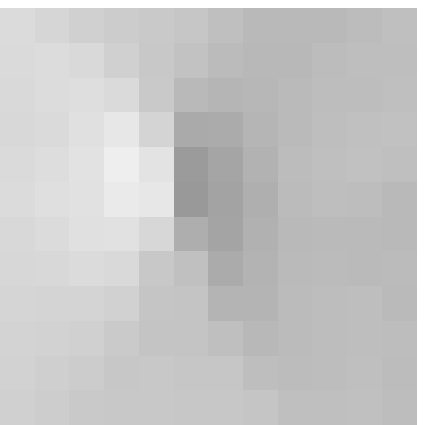,height=2.0cm}
 \psfig{figure=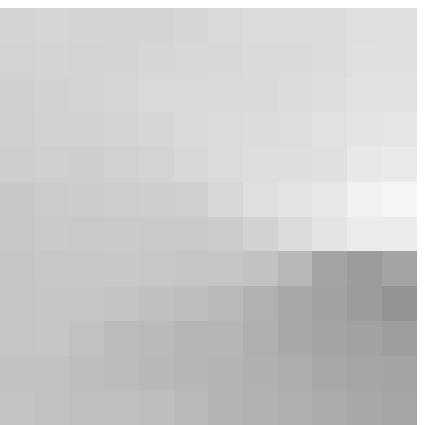,height=2.0cm}
 \psfig{figure=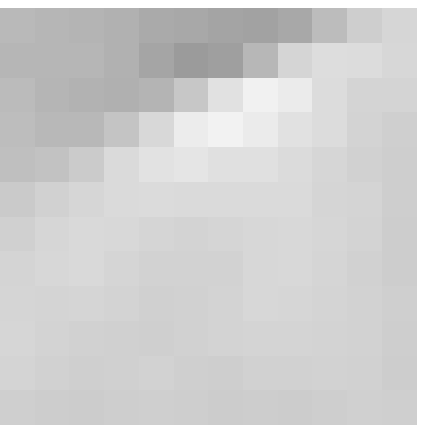,height=2.0cm}
 \psfig{figure=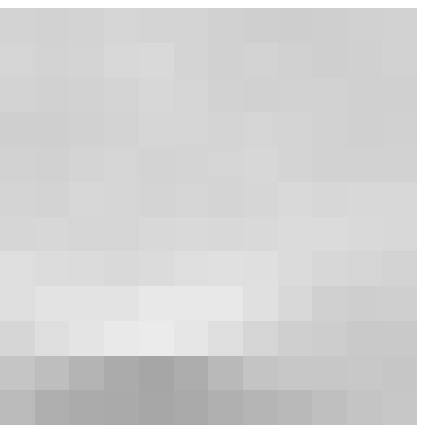,height=2.0cm}
 \psfig{figure=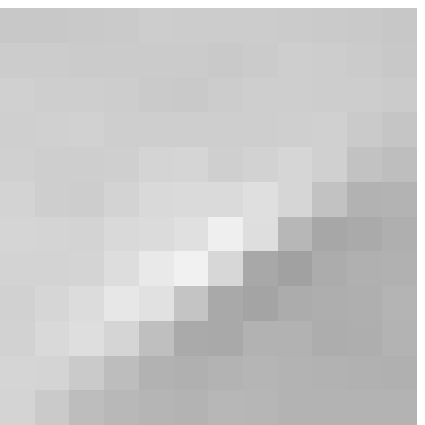,height=2.0cm}
 }}
 \end{center}
 ~\hspace{2.4cm} Bas~ft 0016 \hspace{0.0cm} Bas~ft 0017 \hspace{0.0cm} Bas~ft 0018
 \hspace{0.0cm} Bas~ft 0019 \hspace{0.0cm} Bas~ft 0020
\end{figure}

\end{document}